\documentclass[aps,prc,twocolumn,showpacs,amsmath,preprintnumbers,amssymb,groupedaddress,superscriptaddress]{revtex4}

\usepackage{graphicx}
\usepackage{dcolumn}
\usepackage{bm}
\usepackage{multirow}
\usepackage{hhline}

\begin{document}


\title{$K(892)^*$ Resonance Production in Au+Au and $p+p$ Collisions\\
at $\sqrt{s_{_{NN}}}$ = 200 GeV at RHIC}

\affiliation{Argonne National Laboratory, Argonne, Illinois 60439}
\affiliation{University of Bern, 3012 Bern, Switzerland}
\affiliation{University of Birmingham, Birmingham, United Kingdom}
\affiliation{Brookhaven National Laboratory, Upton, New York
11973} \affiliation{California Institute of Technology, Pasadena,
California 91125} \affiliation{University of California, Berkeley,
California 94720} \affiliation{University of California, Davis,
California 95616} \affiliation{University of California, Los
Angeles, California 90095} \affiliation{Carnegie Mellon
University, Pittsburgh, Pennsylvania 15213} \affiliation{Creighton
University, Omaha, Nebraska 68178} \affiliation{Nuclear Physics
Institute AS CR, 250 68 \v{R}e\v{z}/Prague, Czech Republic}
\affiliation{Laboratory for High Energy (JINR), Dubna, Russia}
\affiliation{Particle Physics Laboratory (JINR), Dubna, Russia}
\affiliation{University of Frankfurt, Frankfurt, Germany}
\affiliation{Institute  of Physics, Bhubaneswar 751005, India}
\affiliation{Indian Institute of Technology, Mumbai, India}
\affiliation{Indiana University, Bloomington, Indiana 47408}
\affiliation{Institut de Recherches Subatomiques, Strasbourg,
France} \affiliation{University of Jammu, Jammu 180001, India}
\affiliation{Kent State University, Kent, Ohio 44242}
\affiliation{Lawrence Berkeley National Laboratory, Berkeley,
California 94720} \affiliation{Massachusetts Institute of
Technology, Cambridge, MA 02139-4307}
\affiliation{Max-Planck-Institut f\"ur Physik, Munich, Germany}
\affiliation{Michigan State University, East Lansing, Michigan
48824} \affiliation{Moscow Engineering Physics Institute, Moscow
Russia} \affiliation{City College of New York, New York City, New
York 10031} \affiliation{NIKHEF, Amsterdam, The Netherlands}
\affiliation{Ohio State University, Columbus, Ohio 43210}
\affiliation{Panjab University, Chandigarh 160014, India}
\affiliation{Pennsylvania State University, University Park,
Pennsylvania 16802} \affiliation{Institute of High Energy Physics,
Protvino, Russia} \affiliation{Purdue University, West Lafayette,
Indiana 47907} \affiliation{University of Rajasthan, Jaipur
302004, India} \affiliation{Rice University, Houston, Texas 77251}
\affiliation{Universidade de Sao Paulo, Sao Paulo, Brazil}
\affiliation{University of Science \& Technology of China, Anhui
230027, China} \affiliation{Shanghai Institute of Applied Physics,
Shanghai 201800, China} \affiliation{SUBATECH, Nantes, France}
\affiliation{Texas A\&M University, College Station, Texas 77843}
\affiliation{University of Texas, Austin, Texas 78712}
\affiliation{Tsinghua University, Beijing 100084, China}
\affiliation{Valparaiso University, Valparaiso, Indiana 46383}
\affiliation{Variable Energy Cyclotron Centre, Kolkata 700064,
India} \affiliation{Warsaw University of Technology, Warsaw,
Poland} \affiliation{University of Washington, Seattle, Washington
98195} \affiliation{Wayne State University, Detroit, Michigan
48201} \affiliation{Institute of Particle Physics, CCNU (HZNU),
Wuhan 430079, China} \affiliation{Yale University, New Haven,
Connecticut 06520} \affiliation{University of Zagreb, Zagreb,
HR-10002, Croatia}

\author{J.~Adams}\affiliation{University of Birmingham, Birmingham, United Kingdom}
\author{M.M.~Aggarwal}\affiliation{Panjab University, Chandigarh 160014, India}
\author{Z.~Ahammed}\affiliation{Variable Energy Cyclotron Centre, Kolkata 700064, India}
\author{J.~Amonett}\affiliation{Kent State University, Kent, Ohio 44242}
\author{B.D.~Anderson}\affiliation{Kent State University, Kent, Ohio 44242}
\author{D.~Arkhipkin}\affiliation{Particle Physics Laboratory (JINR), Dubna, Russia}
\author{G.S.~Averichev}\affiliation{Laboratory for High Energy (JINR), Dubna, Russia}
\author{S.K.~Badyal}\affiliation{University of Jammu, Jammu 180001, India}
\author{Y.~Bai}\affiliation{NIKHEF, Amsterdam, The Netherlands}
\author{J.~Balewski}\affiliation{Indiana University, Bloomington, Indiana 47408}
% ...

\collaboration{STAR Collaboration}\noaffiliation
\date{\today}
\begin{abstract}
    The short-lived $K(892)^{*}$ resonance provides an efficient tool
    to probe properties of the hot and dense
    medium produced in relativistic heavy-ion collisions. We report
    measurements of $K^{*}$ in $\sqrt{s_{_{NN}}}$ = 200 GeV Au+Au
    and $p+p$ collisions reconstructed via its hadronic decay channels
    $K(892)^{*0} \rightarrow K\pi$ and $K(892)^{*\pm} \rightarrow
    K_S^0\pi^{\pm}$ using the STAR detector at RHIC. The $K^{*0}$ mass
    has been studied as a function of $p_T$ in minimum bias $p+p$ and
    central Au+Au collisions. The $K^{*}$ $p_T$ spectra for minimum
    bias $p+p$ interactions and for Au+Au collisions in different
    centralities are presented. The $K^{*}/K$ yield ratios for all centralities
    in Au+Au collisions are found to be significantly lower than the ratio in
    minimum bias $p+p$ collisions, indicating the importance of hadronic
    interactions between chemical and kinetic freeze-outs. A significant
    non-zero $K^{*0}$ elliptic flow ($v_2$) is observed in Au+Au collisions
    and compared to the $K_S^0$ and $\Lambda$ $v_2$. The nuclear
    modification factor of $K^{*}$ at intermediate $p_{T}$ is similar to
    that of $K_{S}^{0}$, but different from $\Lambda$. This establishes a
    baryon-meson effect over a mass effect in the particle production
    at intermediate $p_T$ ($2 < p_T \leq 4$ GeV/$c$).
\end{abstract}
\maketitle

\section{Introduction}
Lattice QCD calculations~\cite{blum} predict a phase transition
from hadronic matter to quark gluon plasma (QGP) at high
temperatures and/or high densities.  Matter under such extreme
conditions can be studied in the laboratory by colliding heavy
nuclei at very high energies. The Relativistic Heavy Ion Collider
(RHIC) at Brookhaven National Laboratory provides collisions of
heavy nuclei and protons at center of mass energies up to
$\sqrt{s_{_{NN}}}$ = 200 GeV. The initial stage of these
collisions can be described as the interpenetration of the nuclei
with partonic interactions at high energy. With the interactions
of the partons in the system, chemical and local thermal
equilibrium of the system may be reached and the QGP may form. As
the system expands and cools down, it will hadronize and
chemically freeze-out. After a period of hadronic interactions,
the system reaches the kinetic freeze-out stage when all hadrons
stop interacting~\cite{rapp1,rapp2,song}. After the kinetic
freeze-out, particles free-stream towards the detectors where our
measurements are performed.

The typical lifetime of a resonance is a few fm/$c$, which is
comparable to the expected lifetime of the hot and dense matter
produced in heavy-ion collisions~\cite{rafelski}. In a hot and
dense system, resonances are in close proximity with other
strongly interacting hadrons. The in-medium effect related to the
high density and/or high temperature of the medium can modify
various resonance properties, such as masses, widths, and even the
mass line shapes~\cite{brown,rapp4,shuryak}. Thus, measurements of
various resonance properties can provide detailed information
about the interaction dynamics in relativistic heavy-ion
collisions~\cite{bielich,rapp3}. Recent measurements~\cite{focus}
by the FOCUS Collaboration for the $K^{*0}$ from charm decays show
that the $K^{*0}$ mass line shape could be changed by the effects
of interference from an $s$-wave and possible other sources.
Distortions of the line shape of $\rho^0$ have also been observed
at RHIC in $p+p$ and peripheral Au+Au collisions~\cite{rho1}.
Dynamical interactions with the surrounding
matter~\cite{rapp4,shuryak,bleicher2}, interference between
various scattering channels~\cite{longacre}, phase space
distortions~\cite{rapp4,shuryak,pbm1,kolb,bron,bleicher2,barz,pratt,granet},
and Bose-Einstein
correlations~\cite{rapp4,shuryak,pbm1,barz,pratt,granet} are
possible explanations for the apparent modification of resonance
properties.

Resonance measurements in the presence of a dense medium can be
significantly affected by two competing effects. Resonances that
decay before kinetic freeze-out may not be reconstructed due to
the rescattering of the daughter particles. In this case, the lost
efficiency in the reconstruction of the parent resonance is
relevant and depends on the time between chemical and kinetic
freeze-outs, the source size, the resonance phase space
distribution, the resonance daughters' hadronic interaction
cross-sections, etc. On the other hand, after chemical freeze-out,
pseudo-elastic interactions~\cite{bleicher0} among hadrons in the
medium may increase the resonance population. This resonance
regeneration depends on the cross-section of the interacting
hadrons in the medium. Thus, the study of resonances can provide
an important probe of the time evolution of the source from
chemical to kinetic freeze-outs and detailed information on
hadronic interactions in the final stage.

In this paper, we study the $K(892)^{*}$ vector meson with a
lifetime of 4 fm/$c$. The kaon and pion daughters of the $K^*$
resonance in the hadronic decay channel $K^* \rightarrow K\pi$ can
interact with other hadrons in the medium. Their rescattering
effect is mainly determined by the pion-pion interaction total
cross section~\cite{proto}, which was measured to be significantly
larger (factor $\sim$5) than the kaon-pion interaction total cross
section~\cite{matison}. The kaon-pion interaction total cross
section determines the regeneration effect that produces the $K^*$
resonance~\cite{bleicher1}. Thus, the final observable $K^*$
yields may decrease compared to the primordial yields, and a
suppression of the $K^*/K$ yield ratio is expected in heavy-ion
collisions. This $K^*$ yield decrease and the $K^*/K$ suppression
compared to elementary collisions, such as $p+p$, at similar
collision energies can be used to roughly estimate the system time
span between chemical and kinetic freeze-outs. Due to the
rescattering of the daughter particles, the low $p_T$ $K^*$
resonances are less likely to escape the hadronic medium before
decaying, compared to high $p_T$ $K^*$ resonances. This could
alter the $K^{*}$ transverse mass ($m_T$) spectra compared to
other particles with similar masses.

The in-medium effects on the resonance production can be
manifested in other observables as well. In a quark coalescence
scenario, the elliptic flow ($v_2$), for non-central Au+Au
collisions, of the $K^*$ resonances produced at chemical
freeze-out might be similar to that of kaons~\cite{lin}. However,
at low $p_T$, the $K^*$ $v_2$ may be modified by the rescattering
effect discussed previously. This rescattering effect also depends
on the hadron distributions in the coordinate space in the system
at the final stage. Thus, a measurement of the $K^*$ $v_2$ at $p_T
\leq 2$ GeV/$c$ compared to the kaon $v_2$ may provide information
on the shape of the fireball in the coordinate space at late
stages.

A study of the relation of the particle production to its
intrinsic properties may reveal its production mechanism. The
nuclear modification factor and $v_2$ have been observed to be
different between $\pi$, $K$ and $p$,
$\Lambda$~\cite{phenixPRC,KsLav2}. In a hydrodynamic limit, the
transverse momentum spectra of produced particles are only
determined by the velocity field and therefore the mass of the
produced particle. In a quark coalescence model, particle
production is related to its quark content. Since stable mesons
($\pi$, $K$) are usually lighter than stable baryons ($p$,
$\Lambda$), the particle type is coupled with the mass. Detailed
studies of $K^*$ (and/or $\phi$) can be of special importance, as
its mass is close to the mass of baryons ($p$, $\Lambda$) but it
is a vector meson. In the intermediate $p_T$ range $2 < p_T \leq
6$ GeV/$c$, identified hadron $v_2$ measurements have shown that
the hadron $v_2$ follows a simple scaling of the number of
constituent quarks in the hadrons: $v_2(p_T) = nv_2^q(p_T/n)$,
where $n$ is the number of constituent quarks of the hadron and
$v_2^q$ is the common elliptic flow for single
quarks~\cite{KsLav2}. Therefore, the $v_2$ for the $K^*$ produced
at hadronization should follow the scaling law with $n=2$.
However, for the $K^*$ regenerated through $K\pi \rightarrow K^*$
in the hadronic stage, $v_2$ should follow the scaling law with
$n=4$~\cite{nonaka}. The measured $K^*$ $v_2$ in the intermediate
$p_T$ region may provide information on the $K^*$ production
mechanism in the hadronic phase and reveal the particle production
dynamics in general.  It is inconclusive whether the difference in
the nuclear modification factor between $K$ and $\Lambda$ is due
to a baryon-meson effect or simply a mass effect~\cite{KsLav2}. We
can use the unique properties of the $K^*$ to distinguish whether
the nuclear modification factor $R_{AA}$ or $R_{CP}$, defined
later in the text (Section V.F.) and in~\cite{highpt}
and~\cite{KsLav2} respectively, in the intermediate $p_T$ region
depends on mass or particle species (i.e. meson/baryon).
Specifically, we can compare the $R_{CP}$ of $K$, $K^*$, and
$\Lambda$, which contain one strange valence quark and are in
groups of ($K$, $K^*$) and $\Lambda$ as mesons vs baryon, or in
groups of $K$ and ($K^*$, $\Lambda$) as different masses.

\section{Experiment}
The data used in this analysis were taken in the second RHIC run
(2001-2002) using the Solenoidal Tracker at RHIC (STAR) with Au+Au
and $p+p$ collisions at $\sqrt{s_{_{NN}}}$ = 200 GeV. The primary
tracking device of the STAR detector is the time projection
chamber (TPC) which is a 4.2 meter long cylinder covering a
pseudo-rapidity range $|\eta| < 1.8$ for tracking with complete
azimuthal coverage ($\Delta\phi = 2\pi$)~\cite{tpc}.

In Au+Au collisions, a minimum bias trigger was defined by
requiring coincidences between two zero degree calorimeters which
are located in the beam directions at $\theta < 2$ mrad and
measure the spectator neutrons. A central trigger corresponding to
the top 10$\%$ of the inelastic hadronic Au+Au cross-section was
defined using both the zero degree calorimeters and the
scintillating central trigger barrel, which surrounds the outer
cylinder of the TPC and triggers on charged particles in the
midpseudorapidity ($|\eta| < $ 0.5) region. In $p+p$ collisions,
the minimum bias trigger was defined using coincidences between
two beam-beam counters that measure the charged particle
multiplicity in forward pseudorapidities (3.3 $ < |\eta| < $ 5.0).

Only events with the primary vertex within $\pm$50 cm from the
center of the TPC along the beam line were selected to insure
uniform acceptance in the $\eta$ range studied. As a result, about
2 $\times$ 10$^6$ top 10\% central Au+Au, 2 $\times$ 10$^6$
minimum bias Au+Au, and 6 $\times$ 10$^6$ minimum bias $p+p$
collision events were used in this analysis. In order to study the
centrality dependence of the $K^*$ production, the events from
minimum bias Au+Au collisions were divided into four centrality
bins from the most central to the most peripheral collisions:
0-10\%, 10-30\%, 30-50\% and 50-80\%, according to the fraction of
the charged hadron reference multiplicity (defined
in~\cite{hadron}) distribution in all events.

\begin{figure}[htp]
\centering
\includegraphics[height=13pc,width=18pc]{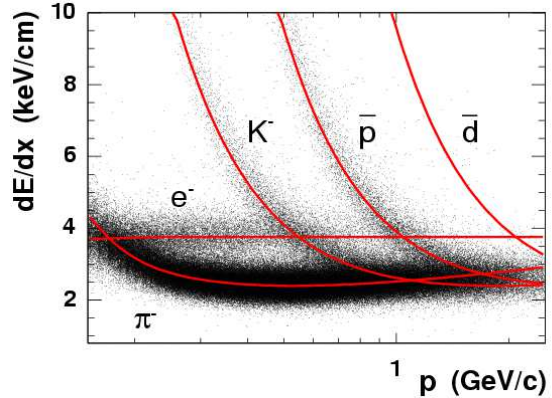}
\caption{(Color online) $dE/dx$ for negative particles vs.
momentum measured by the TPC in Au+Au collisions. The curves are
the Bethe-Bloch parametrization~\cite{pdg} for different particle
species.}\label{fig:dEdx}
\end{figure}

In addition to momentum information, the TPC provides particle
identification for charged particles by measuring their ionization
energy loss ($dE/dx$). The TPC measurement of $dE/dx$ as a
function of the momentum ($p$) is shown in Fig.~\ref{fig:dEdx}.
Different bands seen in Fig.~\ref{fig:dEdx} represent Bethe-Bloch
distributions~\cite{pdg} folded with the experimental resolutions
and correspond to different particle species. Charged pions and
kaons can be identified with their momenta up to about 0.75
GeV/$c$ while protons and anti-protons can be identified with
momenta up to about 1.1 GeV/$c$. In order to quantitatively
describe the particle identification, the variable $N_{\sigma\pi}$
(e.g. pions) was defined as
\begin{equation}
N_{\sigma\pi}=\frac{1}{R}\log\frac{dE/dx_{\text{measured}}}{\langle
dE/dx\rangle_\pi}, \label{eq:NSigma}
\end{equation}
where $dE/dx_\text{measured}$ is the measured energy loss for a
track, $\langle dE/dx\rangle_\pi$ is the expected mean energy loss
for charged pions with a given momentum, and $R$ is the $dE/dx$
resolution which varies between 6\% and 10\% from $p+p$ to central
Au+Au events and depends on the characteristics of each track,
such as the number of $dE/dx$ hits for a track measured in the
TPC, the pseudorapidity of a track, etc. We construct $N_{\sigma
K}$ in a similar way for the charged kaon identification. Specific
analysis cuts (described later) were then applied on
$N_{\sigma\pi}$ and $N_{\sigma K}$ in order to quantitatively
select the charged pion and kaon candidate tracks.

\section{Particle Selections}
In this analysis, the hadronic decay channels of $K(892)^{*0}
\rightarrow K^+\pi^-$, $\overline{K(892)^{*0}} \rightarrow
K^-\pi^+$ and $K(892)^{*\pm} \rightarrow K_S^0\pi^\pm$ were
measured. In the following, the term $K^{*0}$ stands for $K^{*0}$
or $\overline{K^{*0}}$, and the term $K^*$ stands for $K^{*0}$,
$\overline{K^{*0}}$ or $K^{*\pm}$, unless otherwise specified.

\begin{table*}
\caption{\label{tab:cuts}List of track cuts for charged kaon and
charged pion and
  topological cuts for neutral kaon used in the $K^{*}$ analysis in Au+Au
  and $p+p$ collisions. $decayLength$ is the decay length, $dcaDaughters$
  is the distance of closest approach between the daughters, $dcaV0PrmVx$ is
  the distance of closest approach between the reconstructed $K_{S}^{0}$ momentum
  vector and the primary interaction vertex, $dcaPosPrmVx$ is the distance of
  closest approach between the positively charged granddaughter and the primary vertex,
  $dcaNegPrmVx$ is the distance of closest approach between the negatively charged
  granddaughter and the primary vertex, $M_{K_S^0}$ is the $K_S^0$ invariant mass in
  GeV/$c^2$, $NFitPnts$ is the number of fit points of a track in the TPC, $NTpcHits$ is the
  number of hits of a track in the TPC, $MaxPnts$ is the number of maximum possible
  points of a track in the TPC, and $DCA$ is the distance of closest approach to the
  primary interaction point.}
\begin{ruledtabular}
\begin{tabular}{ccccc}
 \multirow{2}{22pt}{Cuts} & \multicolumn{2}{c}{$K^{*0}$} &
    \multicolumn{2}{c}{$K^{*\pm}$}\\ & Au+Au &
    $p+p$ & Daughter $\pi^{\pm}$ & $K_{S}^{0}$\\ \hline
$N_{\sigma K}$
    & (-2.0, 2.0) & (-2.0, 2.0) & & $decayLength > $ 2.0 cm \\
    $N_{\sigma\pi}$ & (-3.0, 3.0) & (-2.0, 2.0) & (-2.0, 2.0) &
    $dcaDaughters < $ 1.0 cm \\ Kaon $p$ (GeV/$c$) & (0.2, 10.0) &
    (0.2, 0.7) & & $dcaV0PrmVx <$ 1.0 cm \\ Kaon $p_{T}$ (GeV/$c$) &
    (0.2, 10.0) & (0.2, 0.7) & & $dcaPosPrmVx >$ 0.5 cm\\ Pion $p$
    (GeV/$c$) & (0.2, 10.0) & (0.2, 10.0) & (0.2, 10.0) &
    $dcaNegPrmVx >$ 0.5 cm\\ Pion $p_{T}$ (GeV/$c$) & (0.2, 10.0) &
    (0.2, 10.0) & (0.2, 10.0) & $M_{K_S^0}$ (GeV/$c^2$): (0.48, 0.51) \\
    $NFitPnts$ & $>$ 15 & $>$ 15 & $>$ 15 & $\pi^{+}$: $NTpcHits >$ 15
    \\ $NFitPnts/MaxPnts$ & $>$ 0.55 & $>$ 0.55 & $>$ 0.55 &
    $\pi^{-}$: $NTpcHits >$ 15 \\ Kaon and pion $\eta$ &
    $|\eta| <$ 0.8 & $|\eta| <$ 0.8 & $|\eta| <$ 0.8 & $\pi^{+}$: $p>$
    0.2 GeV/$c$\\ $DCA$ (cm) & $<$ 3.0 & $<$ 3.0 & $<$ 3.0 &
    $\pi^{-}$: $p >$ 0.2 GeV/$c$ \\ \hline Pair ($K\pi$) $y$ & \multicolumn{4}{c}{$|y| <$ 0.5} \\
    \end{tabular}
\end{ruledtabular}
\end{table*}

Since the $K^*$ decays in such short time that the daughters seem
to originate from the interaction point, only charged kaon and
charged pion candidates whose distance of closest approach to the
primary interaction vertex was less than 3 cm were selected. Such
candidate tracks are defined as ``primary tracks". The charged
$K^*$ first undergoes a strong decay to produce a $K_{S}^{0}$ and
a charged pion herein labeled as the $K^{*\pm}$ daughter pion.
Then, the produced $K_{S}^{0}$ decays weakly into $\pi^+\pi^-$
with $c\tau$ = 2.67 cm. Two oppositely charged pions from the
$K_S^0$ decay are called as the $K^{*\pm}$ granddaughter pions.
The charged daughter pion candidates were selected from primary
track samples and the $K_{S}^{0}$ candidates were selected through
their decay topology.

In Au+Au collisions, charged kaon candidates were selected by
requiring $|N_{\sigma K}| < 2$ while a looser cut $|N_{\sigma\pi}|
< 3$ was applied to select the charged pion candidates to maximize
the statistics for the $K^{*0}$ analysis. Such $N_{\sigma}$ cuts
can only ambiguously select the kaons and pions if applied to the
tracks with their momenta beyond the momentum range specified
earlier. However, these cuts help to significantly reduce the
background. In order to avoid the acceptance drop in the high
$\eta$ range, all kaon and pion candidates were required to have
$|\eta| <$ 0.8. Kaon and pion candidates were also required to
have at least 15 fit points (number of measured TPC hits used in
track fit, maximum 45 fit points) to assure track fitting quality
and good $dE/dx$ resolution. For all the track candidates, the
ratio between the number of TPC track fit points over the maximum
possible points was required to be greater than 0.55 to avoid
selecting split tracks. To maintain reasonable momentum
resolution, only tracks with $p_T$ larger than 0.2 GeV/$c$ were
selected.

In $p+p$ collisions, enough data were available to precisely
measure the $K^{*0}$ mass, width, and invariant yield as a
function of $p_T$. As statistics was not an issue for this
analysis, only kaon candidates with $p<0.7$ GeV/$c$ were used to
ensure clean identification. This kaon momentum cut helped
minimize contamination from misidentified correlated pairs and
thus reduce the systematic uncertainty. In the case of the pion
candidates, the same $p$ and $p_T$ cuts as used in Au+Au
collisions were applied. Charged kaon and pion candidates were
selected by requiring $|N_{\sigma\pi,K}|<2$ to reduce the residual
background. All other track cuts for both kaon and pion candidates
were the same as for Au+Au data.

\begin{figure}[htp]
\centering
\includegraphics[height=13pc,width=18pc]{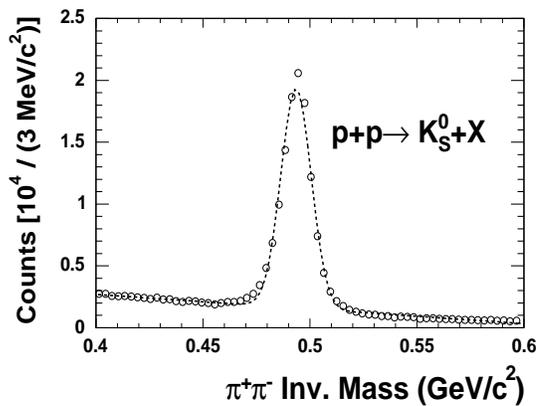}
\caption{$K_S^0$ signal observed in the $\pi^+\pi^-$ invariant
mass distribution reconstructed from the decay topology method via
$K_S^0\rightarrow \pi^+\pi^-$ in $p+p$ collisions. The dashed
curve depicts the Gaussian fit function plus a linear function
representing the background.}\label{fig:kshort}
\end{figure}

The $K^{*\pm}$ was measured only in minimum bias $p+p$
interactions and in peripheral 50-80\% Au+Au collisions. Daughter
pions for the $K^{*\pm}$ reconstruction were required to originate
from the interaction point and pass the same cuts as used for the
$K^{*0}$ analysis in $p+p$ collisions. The $K_{S}^{0}$ was
reconstructed by the decay topology method~\cite{kshort,lambda1}.
The granddaughter charged pion candidates were selected from
global tracks (tracks do not necessarily originate from the
primary collision vertex) with a distance of closest approach to
the interaction point greater than 0.5 cm. Candidates for the
granddaughter charged pions were also required to have at least 15
hit points in the TPC with $p > $ 0.2 GeV/$c$. Oppositely charged
candidates were then paired to form neutral decay vertices. The
distance of closest approach for each pair was required to be less
than 1.0 cm and the neutral decay vertices were required to be at
least 2.0 cm away from the primary vertex to reduce the
combinatorial background. The reconstructed $K_{S}^{0}$ momentum
vector was required to point back to the primary interaction point
within 1.0 cm. Only the $K_{S}^{0}$ candidates with
$\pi^{+}\pi^{-}$ invariant mass between 0.48 and 0.51 GeV/$c^{2}$
were selected. When the $K_S^0$ candidate was paired with the
daughter pion to reconstruct the charged $K^{*}$, tracks were
checked to avoid double-counting among the three tracks used.
Fig.~\ref{fig:kshort} shows the $K_S^0$ signal observed in the
$\pi^+\pi^-$ invariant mass distribution in $p+p$ collisions. The
Gaussian width of the above $K_S^0$ signal is around 7 MeV/$c^2$
which is mainly determined by the momentum resolution of the
detector. Due to detector effects, such as the daughter tracks'
energy loss in the TPC, etc., the $K_S^0$ mass is shifted by $-$3
MeV/c$^2$. The measured $K_S^0$ mass and width agree well with
Monte Carlo ($MC$) simulations, which included the finite momentum
resolution of the detector and the daughter tracks' energy loss in
the TPC.

The $K\pi$ pairs with their parent rapidity ($y$) of $|y|<0.5$
were selected. All the cuts used in this $K^{*}$ analysis are
summarized in Table~\ref{tab:cuts}. After all the above mentioned
cuts have been applied, the $K^*$ reconstruction efficiencies
multiplied by the detector acceptance are shown in
Fig.~\ref{fig:efficiency}.

\section{Extraction of the $K^*$ Signal}
In Au+Au collisions, up to several thousand charged tracks per
event originate from the primary collision vertex. The daughters
from $K^*$ decays are topologically indistinguishable from other
primary particles.  The measurement was performed by calculating
the invariant mass for each $K\pi$ pair in an event. The
$K^{\pm}\pi^{\mp}$ invariant mass distribution is shown in
Fig.~\ref{fig:mixevent} as open circles. The unlike-sign $K\pi$
invariant mass distribution derived in this manner was mostly from
random $K\pi$ combinatorial pairs. The signal to background is
between 1/200 for minimum bias Au+Au and 1/10 for minimum bias
$p+p$. The overwhelming combinatorial background distribution can
be obtained and subtracted from the unlike-sign $K\pi$ invariant
mass distribution in two ways:
\begin{itemize}
\item the mixed-event technique: reference background
distribution is built with uncorrelated unlike-sign kaons and
pions from different events;
\item the like-sign technique: reference background distribution is made from like-sign kaons and pions
in the same event.
\end{itemize}

The mixed-event technique has been successfully used in the
measurement of resonances at RHIC, such as the $K(892)^{*0}$ in
Au+Au collisions at $\sqrt{s_{_{NN}}} = $ 130 GeV~\cite{kstar130}
and the $\phi$ in Au+Au collisions at $\sqrt{s_{_{NN}}} = $ 130
and 200 GeV~\cite{phi1,phi2}. This technique was also used in the
measurement of $\Lambda$ production in Au+Au collisions at
$\sqrt{s_{_{NN}}}$ = 130 GeV, and the results agree well with
those from the decay topology method~\cite{lambda1,lambda2}. The
like-sign technique has been successfully applied in measuring
$\rho(770)^0 \rightarrow \pi^+\pi^-$ production in $p+p$ and
peripheral Au+Au collisions at $\sqrt{s_{_{NN}}} = $ 200 GeV at
RHIC~\cite{rho1}.

\subsection{Mixed-Event Technique}
In order to subtract the uncorrelated pairs from the unlike-sign
$K\pi$ invariant mass distribution obtained from the same events,
an unlike-sign $K\pi$ invariant mass spectrum from mixed events
was obtained. In order to keep the event characteristics as
similar as possible among different events, the whole data sample
was divided into 10 bins in charged particle multiplicity and 10
bins in the collision vertex position along the beam direction.
Only pairs from events in the same multiplicity and vertex
position bins were selected.

\begin{figure}[htp]
\centering
\includegraphics[height=14pc,width=18pc]{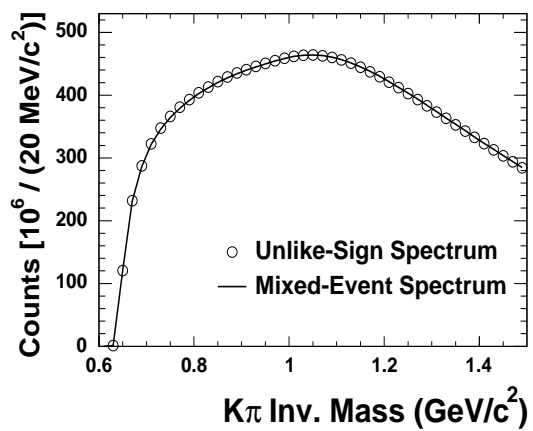}
\caption{The unlike-sign $K\pi$ invariant mass distribution (open
symbols) and the mixed-event $K\pi$ invariant mass distribution
after normalization (solid curve) from minimum bias Au+Au
collisions.} \label{fig:mixevent}
\end{figure}

In the unlike-sign invariant mass distribution from an event,
$K^{+}_1\pi^{-}_1$ and $K^{-}_1\pi^{+}_1$ pairs were sampled which
include the desired $K^*$ signal and the background. In the
mixed-event spectrum, $K^{+}_1\pi^{-}_i$ , $K^{-}_1\pi^{+}_i$,
$K^{+}_i\pi^{-}_1$, and $K^{-}_i\pi^{+}_1$ pairs were sampled for
the background estimation. The subscripts 1 and $i$ correspond to
event numbers with $i\neq$1. The number of events to be mixed was
chosen to be 5, so that the total number of entries in the
mixed-event invariant mass distribution was $\sim$10 times that of
the total number of entries in the distribution from the same
events. Thus the mixed-event spectrum needs to be normalized in
order to subtract the background in the unlike-sign spectrum.
Since the $K\pi$ pairs with invariant mass greater than 1.1
GeV/$c^2$ are less likely to be correlated in the unlike-sign
distribution, the normalization factor was calculated by taking
the ratio between the number of entries in the unlike-sign and the
mixed-event distributions for invariant mass greater than 1.1
GeV/$c^{2}$. The solid curve in Fig.~\ref{fig:mixevent}
corresponds to the mixed-event $K\pi$ pair invariant mass
distribution after normalization. The mixed-event distribution was
then subtracted from the unlike-sign distribution as follows:
\begin{eqnarray}
N_{K^{*0}}(m)=N_{K^{+}_1\pi^{-}_1}(m)+N_{K^{-}_1\pi^{+}_1}(m)\nonumber\\
-R\times\sum_{i=2}^{6}[N_{K^{+}_1\pi^{-}_i}(m)+N_{K^{-}_1\pi^{+}_i}(m)\nonumber\\
+N_{K^{+}_i\pi^{-}_1}(m)+N_{K^{-}_i\pi^{+}_1}(m)],
\label{eq:mixed_subtraction}
\end{eqnarray}
where $N$ is the number of entries in a bin with its center at the
$K\pi$ pair invariant mass $m$ and $R$ is the normalization
factor. After the mixed-event background subtraction, the $K^{*0}$
signal is visible as depicted by the open star symbols in
Fig.~\ref{fig:background}.

\subsection{Like-Sign Technique}
The like-sign technique is another approach to subtract the
background of non-correlated pairs from the unlike-sign $K\pi$
invariant mass distribution from the same events. The uncorrelated
background in the unlike-sign $K\pi$ distribution was described by
using the invariant mass distributions obtained from uncorrelated
$K^+ \pi^+$ and $K^- \pi^-$ pairs from the same events.

\begin{figure}[htp]
\centering
\includegraphics[height=14pc,width=18pc]{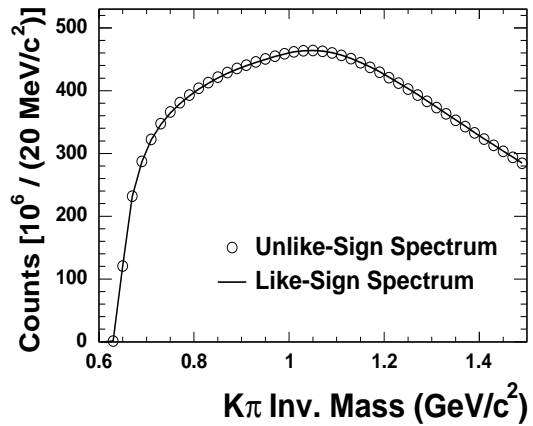}
\caption{The unlike-sign $K\pi$ invariant mass distribution (open
symbols) and the like-sign $K\pi$ invariant mass distribution
(solid curve) from minimum bias Au+Au
collisions.}\label{fig:likesign}
\end{figure}

In the unlike-sign $K\pi$ invariant mass spectrum,
$K^{+}_1\pi^{-}_1$ and $K^{-}_1\pi^{+}_1$ pairs were sampled.
$K^{+}_1\pi^{+}_1$ and $K^{-}_1\pi^{-}_1$ pairs were sampled in
the like-sign $K\pi$ invariant mass distribution. Since the number
of positive and negative particles may not be the same in
relativistic heavy-ion collisions, in order to correctly subtract
the subset of non-correlated pairs in the unlike-sign $K\pi$
distribution, the like-sign $K\pi$ invariant mass distribution was
calculated as follows:
\begin{equation}
N_{\text{Like-Sign}}(m)=2\times\sqrt{N_{K^{+}_1\pi^{+}_1}(m)\times
N_{K^{-}_1\pi^{-}_1}(m)},
\label{eq:likesign}
\end{equation}
where $N$ is the number of entries in a bin with its center at the
$K\pi$ pair invariant mass $m$. The unlike-sign and the like-sign
invariant mass distributions are shown in Fig.~\ref{fig:likesign}.
The like-sign spectrum was then subtracted from the unlike-sign
distribution:
\begin{equation}
N_{K^{*0}}(m)=N_{K^{+}_1\pi^{-}_1}(m)+N_{K^{-}_1\pi^{+}_1}(m)-N_{\text{Like-Sign}}(m).
\label{eq:likesign_subtraction}
\end{equation}
The like-sign background subtracted $K\pi$ invariant mass
distribution corresponds to the solid square symbols in
Fig.~\ref{fig:background}, where the $K^{*0}$ signal is now
visible.

Compared to the mixed-event technique, the like-sign technique has
the advantage that the unlike-sign and like-sign pairs are taken
from the same event, so there is no event structure difference
between the two distributions due to effects such as elliptic
flow. The short-coming of this technique is that the like-sign
distribution has larger statistical uncertainties compared to the
mixed-event spectrum, since the statistics in the mixed-event and
like-sign techniques are driven by the number of events mixed and
the number of kaons and pions produced per event,
respectively~\cite{haibin}. Therefore, in this analysis, the
mixed-event technique was used to reconstruct the $K^{*}$ signal
whereas the like-sign technique was used to study the sources of
the residual background under the $K^{*0}$ peak after mixed-event
background subtraction as discussed in details in the following
text.

\subsection{Describing the Residual Background}
\begin{figure}[htp]
\centering
\includegraphics[height=14pc,width=18pc]{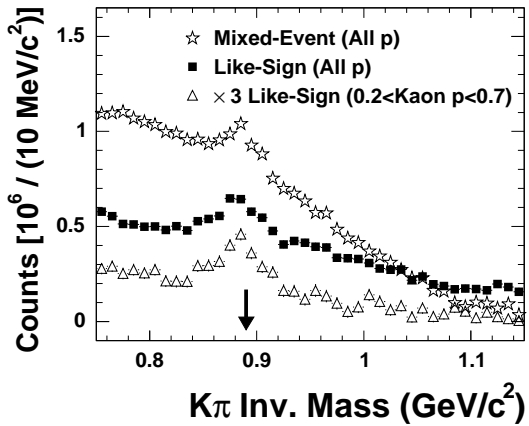}
\caption{The $K\pi$ invariant mass distributions after
event-mixing background subtraction (open star symbols) and
like-sign background subtraction with different daughter momentum
cuts (0.2 $<$ Kaon and Pion $p <$ 10 GeV/$c$ for filled square
symbols, 0.2 $<$ Kaon $p <$ 0.7 GeV/$c$ and 0.2 $<$ Pion $p <$ 10
GeV/$c$ for open triangle symbols) demonstrating the sources of
the residual background in minimum bias Au+Au collisions. The open
triangle symbols have been scaled up by a factor of 3 in order to
increase the visibility. The arrow depicts the standard $K^{*0}$
mass of 896.1 MeV/$c^2$~\cite{pdg}.}\label{fig:background}
\end{figure}

The unlike-sign $K\pi$ invariant mass distribution after
mixed-event background subtraction is represented by the open star
symbols in Fig.~\ref{fig:background}, where the $K^{*0}$ signal is
clearly observed. The mixed-event technique removes only the
uncorrelated background pairs in the unlike-sign spectrum. As a
consequence, residual correlations near the $K^{*0}$ mass range
were not subtracted by the mixed-event spectrum. This residual
background may come from three dominant sources:
\begin{itemize}
\item elliptic flow in non-central Au+Au collisions;
\item correlated real $K\pi$ pairs;
\item correlated but misidentified pairs.
\end{itemize}

The overlapping region of non-central Au+Au collisions has an
elliptic shape in the plane perpendicular to the beam axis. Each
non-central Au+Au event has a unique reaction plane angle. The
azimuthal distributions for kaons and pions may be different for
different events. Thus, the unlike-sign $K\pi$ pair invariant mass
spectrum may have a different structure than the mixed-event
invariant mass distribution. This structural difference may lead
to a significant residual background in the unlike-sign $K\pi$
invariant mass spectrum after mixed-event background
subtraction~\cite{gaudichet}.

In the like-sign technique, the unlike-sign $K\pi$ spectrum and
the like-sign distribution are obtained from the same events.
Therefore, no correlations due to elliptic flow should be present
in the unlike-sign $K\pi$ invariant mass spectrum after like-sign
background subtraction. In Fig.~\ref{fig:background}, the solid
square symbols represent the unlike-sign $K\pi$ invariant mass
distribution after like-sign background subtraction. The amplitude
of the residual background below the peak after the like-sign
background subtraction is about a factor of 2 smaller than after
the mixed-event background subtraction, while the amplitude of the
$K^{*0}$ signal remains the same. This indicates that part of the
residual background in the spectrum after mixed-event background
subtraction was induced by elliptic azimuthal anisotropy.

In the $K^{*0}$ analysis in Au+Au collisions, since the kaons and
pions are selected with 0.2 $< p <$ 10.0 GeV/$c$, a pion (kaon)
with $p>$ 0.75 GeV/$c$ may be misidentified as a kaon (pion). A
proton with $p >$ 1.1 GeV/$c$ may be misidentified as either a
kaon or a pion, or both, depending on whether kaons or pions are
being selected. Electrons and positrons which cross the kaon
(pion) band in the $dE/dx$ plot shown in Fig.~\ref{fig:dEdx} may
be misidentifed as kaons (pions). Thus, the daughters from
$\rho^{0}\rightarrow\pi^+\pi^-$, $\phi\rightarrow K^+ K^-$,
$\Lambda\rightarrow \pi^- p$, etc. could be falsely identified as
a $K\pi$ pair if the daughter momenta are beyond the particle
identification range. The invariant mass calculated from these
misidentified pairs cannot be subtracted away by the mixed-event
background and remains as part of the residual background.

In Fig.~\ref{fig:background}, the open triangle symbols correspond
to the unlike-sign $K\pi$ spectrum after like-sign background
subtraction with 0.2 $< p <$ 0.7 GeV/$c$ and 0.2 $< p <$ 10.0
GeV/$c$ for the kaon and the pion, respectively. These momentum
cuts allow only correlated $K\pi$ real pairs and pairs in which a
kaon or a proton was misidentified as a pion to contribute to the
background subtracted spectrum. Compared to the solid square
symbols in Fig.~\ref{fig:background}, the residual background
represented by the open triangle symbols is reduced by a factor of
6 and the $K^{*0}$ signal is a factor of 2 smaller. This indicates
that particle misidentification of the decay products of $\rho$,
$\omega$, $\eta$, $K_S^0$, $\Lambda$, etc. indeed causes false
correlations to appear in the background subtracted distribution.

Correlated real $K\pi$ pairs from real particle decays, such as
higher mass resonant states in the $K-\pi$ system and particle
decay modes with three or more daughters where two of them are a
$K\pi$ pair, as well as the nonresonant $K-\pi$ $s$-wave
correlation also contribute to the unlike-sign $K\pi$ spectrum.
These correlated $K\pi$ pairs contribute to the residual
background, since they are not present in the like-sign and
mixed-event distributions. There is no efficient cut to remove
these real correlations from the residual background.

\section{Analysis and Results}

\subsection{$K^{*0}$ Mass and Width}

\begin{figure}[htp]
\centering
\includegraphics[height=22pc,width=18pc]{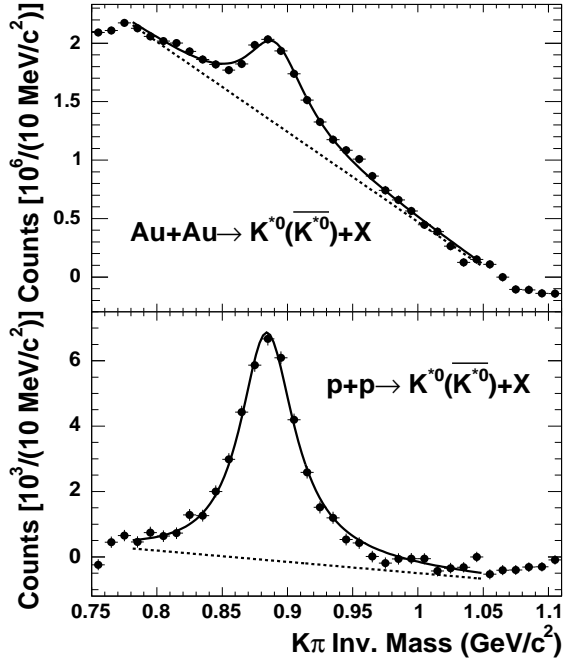}
\caption{The $K\pi$ invariant mass distribution integrated over
the $K^*$ $p_T$ for central Au+Au (upper panel) and minimum bias
$p+p$ (lower panel) interactions after the mixed-event background
subtraction. The solid curves are the fits to
Eq.~\ref{eq:fit-function} with $T_{fo}$ = 120 MeV and $p_T$ = 1.8
GeV/$c$ for central Au+Au and $T_{fo}$ = 160 MeV and $p_T$ = 0.8
GeV/$c$ for $p+p$, respectively. The dashed lines are the linear
function representing the residual
background.}\label{fig:NeutralAuAupp}
\end{figure}

Figure~\ref{fig:NeutralAuAupp} depicts the mixed-event background
subtracted $K\pi$ invariant mass distributions ($M_{K\pi}$)
integrated over the $K^*$ $p_T$ for central Au+Au (upper panel)
and for minimum bias $p+p$ (lower panel) interactions. The mass of
the $K^{*0}$ was fit to the function:
\begin{eqnarray}
BW \times PS + RBG, \label{eq:fit-function}
\end{eqnarray}
where $BW$ is the relativistic $p$-wave Breit-Wigner function
\cite{upc}:
\begin{eqnarray}
BW = \frac{M_{K\pi} \Gamma M_0}{(M_{K\pi}^2-M_0^2)^2 + M_0^2
\Gamma^2}, \label{eq:breit-wigner}
\end{eqnarray}
$PS$ is the Boltzmann factor \cite{rapp4,shuryak,pbm1,kolb}:
\begin{eqnarray}
PS = \frac{M_{K\pi}}{\sqrt{M_{K\pi}^2 + p_T^2}} \times
\exp\left(-\frac{\sqrt{M_{K\pi}^2 + p_T^2}}{T_{fo}}\right)
\label{eq:phase-space}
\end{eqnarray}
that accounts for phase space, and $RBG$ is the linear function:
\begin{eqnarray}
RBG = a + b M_{K\pi} \label{eq:background}.
\end{eqnarray}
that represents the residual background. Within this
parametrization, $T_{fo}$ is the temperature at which the
resonance is emitted \cite{shuryak} and
\begin{equation}
\Gamma = \frac{\Gamma_0 M_0^4}{M_{K\pi}^4} \times \left[
\frac{(M_{K\pi}^2-M_\pi^2-M_K^2)^2-4M_\pi^2
M_K^2}{(M_0^2-M_\pi^2-M_K^2)^2-4M_\pi^2 M_K^2} \right]^{3/2}
\label{eq:gamma}
\end{equation}
is the momentum dependent width \cite{upc}. In addition, $M_0$ is
the $K^{*}$ mass, $\Gamma_0$ is the $K^{*}$ width, $p_T$ is the
$K^{*}$ transverse momentum, $M_\pi$ is the pion mass, and $M_K$
is the kaon mass.

The $PS$ factor accounts for $K^{*}$ produced through kaon and
pion scattering, or $K+\pi\rightarrow K^{*}\rightarrow K+\pi$. In
Au+Au collisions, the thermal freeze-out temperature $T_{fo}$ = 90
MeV was measured at STAR~\cite{meanpt}. However, resonances can be
produced over a range of temperature inside the hadronic system
and not all resonances are emitted at the point where the system
freezes out at $T_{fo}$ = 90 MeV. As a result, the temperature
chosen in the PS factor was 120 MeV according to~\cite{shuryak}.
The temperature of $T_{fo}$ = 90 MeV was also used to estimate the
systematic uncertainties which are about 1.5 MeV/$c^2$ for masses
and 5 MeV/$c^2$ due the choice of $T_{fo}$. In $p+p$ collisions,
particle production is well reproduced by the statistical
model~\cite{becattini} with $T_{fo}$ = 160 MeV and therefore this
was the temperature used in the $PS$ factor. $p_T$ = 1.8 GeV/$c$
and 0.8 GeV/$c$ were chosen in the $PS$ factor for the Au+Au and
$p+p$ collisions respectively which are the centers of the entire
measured $p_T$ ranges (0.4 $<p_T<$ 3.2 GeV/$c$ for Au+Au and
$p_T<$1.6 GeV/$c$ for p+p).

\begin{figure}[htp]
\centering
\includegraphics[height=22pc,width=18pc]{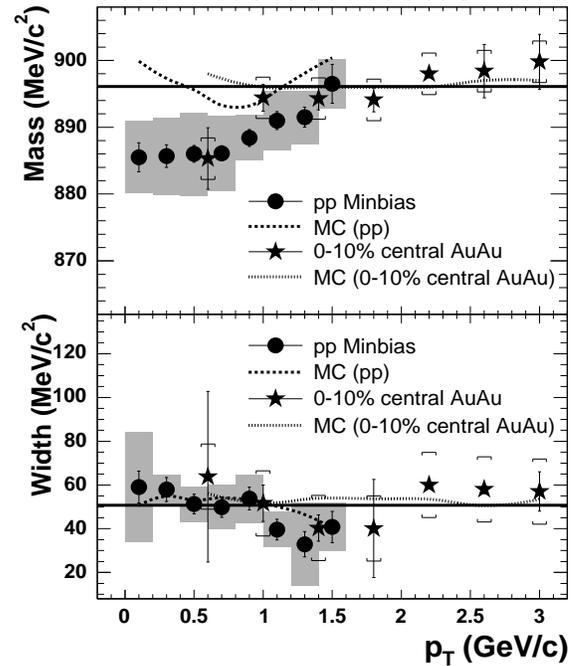}
\caption{The $K^{*0}$ mass (upper panel) and width (lower panel)
as a function of $p_T$ for minimum bias $p+p$ interactions and for
central Au+Au collisions. The solid straight lines are the
standard $K^{*0}$ mass (896.1 MeV/$c^2$) and width (50.7
MeV/c$^2$)~\cite{pdg}, respectively. The dashed and dotted curves
are the $MC$ results in minimum bias $p+p$ and for central Au+Au
collisions, respectively, after considering detector effects and
kinematic cuts. The grey shadows (caps) indicate the systematic
uncertainties for the measurement in minimum bias $p+p$
interactions (central Au+Au collisions).}\label{fig:MassWidth}
\end{figure}

Mixed-event background subtracted $K\pi$ invariant mass
distributions were obtained for different $p_T$ bins, and each
$p_T$ bin was fit to Eq.~\ref{eq:fit-function} with the $K^{*0}$
mass, width, and uncorrected yield as free parameters. The
$\chi^2/ndf$ of the fit varies between 0.6 and 1.7 for all $p_T$
bins except for two $p_T$ bins (3.8 for the $2.0<p_T<2.4$ GeV/$c$
bin and 2.6 for the $2.4<p_T<2.8$ GeV/$c$ bin) in the central
Au+Au data, where the uncertainties of the mass and width values
are not well constrained. Figure~\ref{fig:MassWidth} shows the
$K^{*0}$ mass (upper panel) and width (lower panel) for central
Au+Au and for minimum bias $p+p$ interactions as a function of the
$K^{*0}$ $p_T$. $MC$ calculations for the $K^{*0}$ mass and width
were obtained by simulating $K^{*0}$ with standard mass and width
values~\cite{pdg} and passing them through the same reconstruction
steps and kinematic cuts as the real data. The results from such
simulations are also depicted in Fig.~\ref{fig:MassWidth}. The
deviations between the $MC$ results and the standard mass and
width values are mainly due to the kinematic cuts (track $p$ and
$p_T$ cuts, etc.). For example, the $p_T>0.2$ GeV/$c$ cut results
in the rise of the mass at low $p_T$ and the kaon $p<0.7$ GeV/$c$
cut in $p+p$ causes the rise of the mass and the drop of the width
at higher $p_T$. Our $MC$ studies indicate that the deviations
induced by kinematic cuts are not sufficient to explain the mass
shift seen in the data.

The systematic uncertainties in the $K^{*0}$ mass and width for
the measurement in minimum bias $p+p$ interactions were evaluated
bin-by-bin by varying the particle types (either $K^{*0}$ or
$\overline{K^{*0}}$ ), the methods in the background subtraction
(mixed-event or like-sign), the residual background functions
(exponential or second order polynomial functions), the dynamical
cuts, the track types (primary tracks or global tracks), and by
considering the detector effects (different TPC magnetic field
directions, different sides of the TPC detector, etc.). Due to the
limited statistics, the systematic uncertainties (3.1 MeV/c$^2$
for masses and 14.9 MeV/c$^2$ for widths) in central Au+Au
interactions were only estimated using the entire measured $p_T$
range ($0.4<p_T<3.2$ GeV/$c$) following the above steps. More
detailed discussions about the systematic uncertainty studies can
be found in~\cite{haibin}. In minimum bias p+p interactions, the
$K^{*0}$ masses at low $p_T$ (first 2-3 data points) are lower
than the $MC$ results at a 2-3$\sigma$ level.

\subsection{$m_T$ and $p_T$ Spectra}

\begin{figure}[htp]
\centering
\includegraphics[height=22pc,width=18pc]{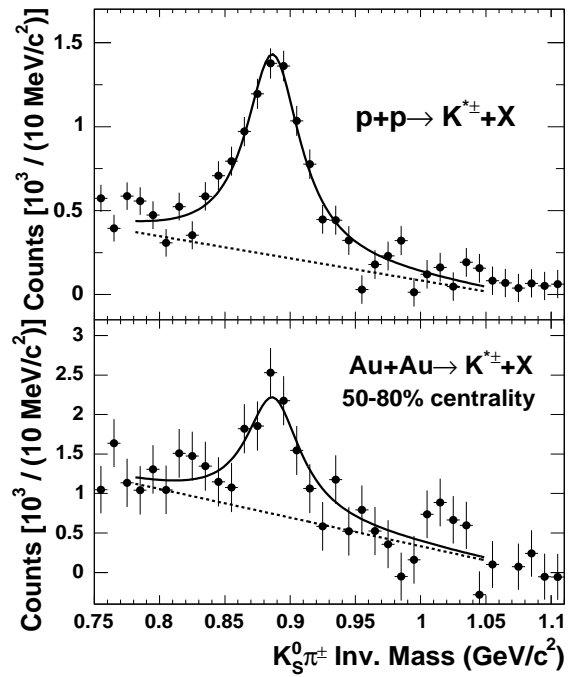}
\caption{The $K_{S}^{0}\pi^{\pm}$ invariant mass distribution
integrated over the $K^{*\pm}$ $p_T$ for minimum bias $p+p$
collisions (upper panel) and for the 50-80\% of the inelastic
hadronic Au+Au cross-section (lower panel) after the mixed-event
background subtraction. The solid curves are fits to
Eq.~\ref{eq:sfit-function} and the dashed lines are the linear
function representing the residual
background.}\label{fig:ChargedppAuAu}
\end{figure}

Mixed-event background subtracted $K\pi$ invariant mass
distributions were obtained for different $p_T$ bins, and each
$p_T$ bin was fit to function:
\begin{eqnarray}
SBW + RBG, \label{eq:sfit-function}
\end{eqnarray}
where $SBW$ is the non-relativistic Breit-Wigner function
\cite{pdg}:
\begin{eqnarray}
SBW = \frac{\Gamma_0}{(M_{K\pi}-M_0)^2+\Gamma_0^2/4}
\label{eq:sbreit-wigner}
\end{eqnarray}
and $RBG$ is the linear function from Eq.~\ref{eq:background} that
represents the residual background. The fit sensitivity to
statistical fluctuations in the $K^*$ raw yield was reduced by
fixing the mass and width in the fit according to the values
obtained from the free parameter fit with the same simplified $BW$
function. The $\chi^2/ndf$ of the fit varies between 0.7 and 1.8
for all $p_T$ bins except for two $p_T$ bins ($\sim$3.0 for the
$2.0<p_T<2.4$ GeV/$c$ bin and $\sim$2.7 for the $2.4<p_T<2.8$
GeV/$c$ bin) in Au+Au data. The $K^{*}$ raw yield was also
obtained by fitting the data to the $BW$ function from
Eq.~\ref{eq:breit-wigner} with all parameters free in the fit. The
difference in the raw yields between the two fit functions was
included in the systematic uncertainties. The $K_{S}^{0}\pi^{\pm}$
invariant mass distribution fit to Eq.~\ref{eq:sfit-function}
after the mixed-event background subtraction is shown in
Fig.~\ref{fig:ChargedppAuAu} for minimum bias $p+p$ collisions
(upper panel) and for the 50-80\% of the inelastic hadronic Au+Au
cross-section (lower panel).

\begin{figure}[htp]
\centering
\includegraphics[height=13pc,width=18pc]{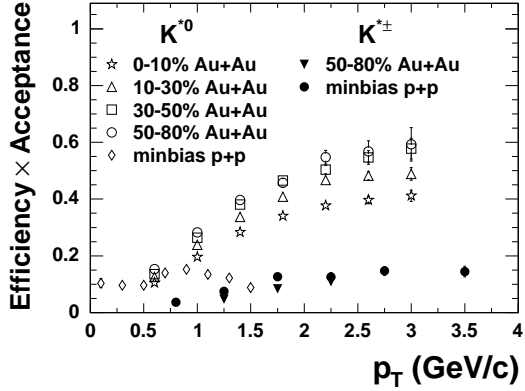}
\caption{The $K^{*0}$ and $K^{*\pm}$ reconstruction efficiency
multiplied by the detector acceptance as a function of $p_T$ in
minimum bias $p+p$ and different centralities in Au+Au
collisions.}\label{fig:efficiency}
\end{figure}

About 6$\times$10$^{6}$, 2$\times$10$^6$ and 5.6$\times$10$^4$
$K^{*0}$ signals were reconstructed from top 10\% central Au+Au,
minimum bias Au+Au and minimum bias $p+p$ collisions respectively
while about 1.2$\times$10$^4$ and 10$^4$ $K^{*\pm}$ were observed
in the 50-80\% Au+Au and minimum bias $p+p$ collisions
respectively. The $K^{*0}$ and $K^{*\pm}$ raw yields obtained for
different $p_T$ bins in Au+Au and minimum bias $p+p$ collisions
were then corrected for the detector acceptance and efficiency
(shown in Fig.~\ref{fig:efficiency}) determined from a detailed
simulation of the TPC response using GEANT \cite{hminus}. The
corresponding branching ratios were also taken into account. In
addition, the yields in $p+p$ were corrected for the collision
vertex finding efficiency of 86\%.

\begin{figure}[htp]
\centering
\includegraphics[height=18pc,width=18pc]{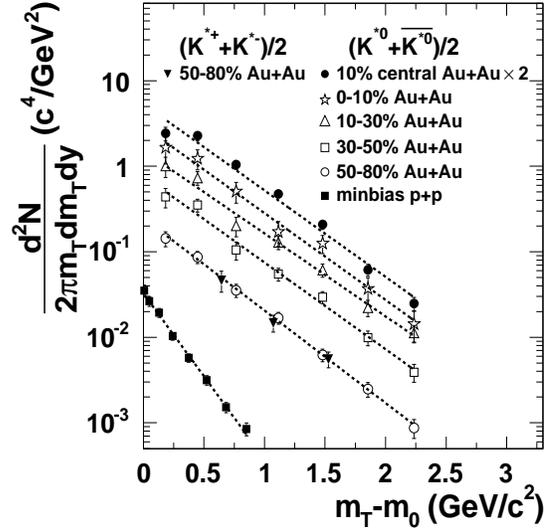}
\caption{The ($K^{*}+\overline{K^{*}}$)/2 invariant yields as a
function of $m_T-m_0$ for $|y|<$ 0.5 from minimum bias $p+p$ and
different centralities in Au+Au collisions. The top 10\% central
data have been multiplied by two for clarity. The lines are fits
to Eq.~\ref{eq:exponential}. The errors shown are the quadratic
sum of statistical and systematic (in the level of 10\%)
uncertainties.}\label{fig:dndy}
\end{figure}

The transverse mass ($m_T$) distributions of the midrapidity
$(K^{*0}+\overline{K^{*0}})/2$ invariant yields in central Au+Au,
four different centralities in minimum bias Au+Au, and minimum
bias $p+p$ collisions are depicted in Fig.~\ref{fig:dndy}. The
$(K^{*+}+K^{*-})/2$ invariant yields for the most peripheral
50-80\% Au+Au collisions are also shown for comparison. The
$K^{*0}$ invariant yield [$d^2N/(2\pi m_Tdydm_T)$] distributions
were fit to an exponential function:
\begin{eqnarray}
\frac{1}{2\pi m_T}\frac{d^2N}{dydm_T}=\frac{dN}{dy}
\frac{1}{2\pi T(m_0+T)}  \nonumber \\
\exp\left(\frac{-(m_T-m_0)}{T}\right), \label{eq:exponential}
\end{eqnarray}
where $dN/dy$ is the $K^{*0}$ yield at $|y| < 0.5$ and $T$ is the
inverse slope parameter. The extracted $dN/dy$ and $T$ parameters
are listed in Table~\ref{tab:dndy}. The systematic uncertainties
on the $K^{*0}$ $dN/dy$ and $T$ in Au+Au and $p+p$ collisions were
estimated by comparing different Breit-Wigner functions, particle
types (either $K^{*0}$ or $\overline{K^{*0}}$), residual
background functions (exponential or second order polynomial
functions), dynamical cuts, and by considering the detector
effects. More detailed discussions about the systematic
uncertainty studies can be found in~\cite{haibin}. The $K^{*0}$
invariant yield increases from $p+p$ collisions to peripheral
Au+Au and to central Au+Au collisions. The inverse slope of the
$K^{*0}$ spectra for all centrality bins of Au+Au collisions is
significantly larger than in minimum bias $p+p$ collisions.

\begin{table}[htp]
\caption{\label{tab:dndy}The $K^{*0}$ $dN/dy$ and $T$ for $|y| < $
0.5 from central Au+Au,
  four different centralities in minimum bias Au+Au, and minimum
  bias $p+p$ collisions. The first error is statistical, the second is systematic.}
\begin{ruledtabular}

\begin{tabular}{ccc}
 &$dN/dy$ & $T$ (MeV) \\ \hline
    top 10\% central  & 10.18$\pm$0.46$\pm$1.88 & 427$\pm$10$\pm$46 \\
    0-10\%            & 10.48$\pm$1.45$\pm$1.94 & 428$\pm$31$\pm$47 \\
    10-30\%         & 5.86$\pm$0.56$\pm$1.08 & 446$\pm$23$\pm$49 \\
    30-50\%         & 2.81$\pm$0.25$\pm$0.52 & 427$\pm$18$\pm$46 \\
    50-80\%         & 0.82$\pm$0.06$\pm$0.15 & 402$\pm$14$\pm$44 \\
    $p+p$               & (5.08$\pm$0.17$\pm0.61$)$\times$10$^{-2}$ & 223$\pm$8$\pm$9 \\
\end{tabular}
\end{ruledtabular}
\end{table}

\begin{figure}[htp] \centering
\includegraphics[height=13pc,width=18pc]{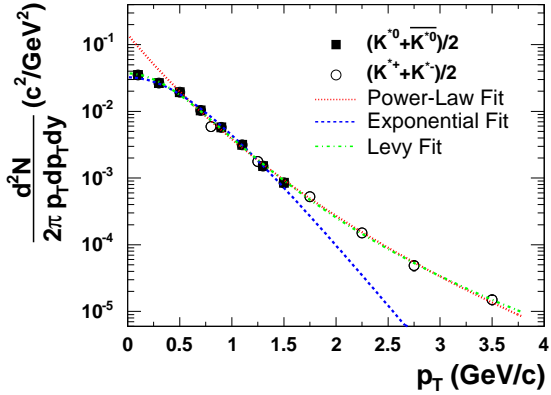}
\caption{(Color online) The invariant yields for both
($K^{*0}+\overline{K^{*0}}$)/2 and ($K^{*+}+K^{*-}$)/2 as a
function of $p_T$ for $|y|<$ 0.5 in minimum bias $p+p$
interactions. The dotted curve is the fit to the power-law
function from Equation~\ref{eq:power-law} for $p_T > 0.5$ GeV/$c$
and extended to lower values of $p_T$. The dashed curve is the
$K^{*0}$ spectrum fit to the exponential function from
Equation~\ref{eq:exponential} and extended to higher values of
$p_T$. The dashed-dotted curve is the fit to the Levy function
from Equation~\ref{eq:levy} for $p_T < 4$ GeV/$c$. Errors are
statistical only.}\label{fig:pT}
\end{figure}

Theoretical calculations~\cite{powerlaw} indicate that in $p+p$
collisions, particle production is dominated by hard processes for
$p_T$ above 1.5 GeV/$c$ while soft processes dominate at low
$p_T$. Thus in the $K^{*}$ $p_T$ spectrum, a power-law shape for
$p_T$ above 1.5 GeV/$c$ and an exponential shape at lower $p_T$
should be expected. In minimum bias $p+p$ collisions, due to the
cut on the kaon daughter of $p < 0.7$ GeV/$c$, only the $K^{*0}$
spectrum for $p_T < 1.6$ GeV/$c$ was measured. As a result, the
$K^{*0}$ $m_T$ spectrum in minimum bias $p+p$ collisions can be
well described by the commonly used exponential function, as shown
in Fig.~\ref{fig:dndy}. The $K^{*}$ $p_T$ spectrum can be extended
to higher $p_T$ by measuring the $K^{*\pm}$ signals.
Figure~\ref{fig:pT} shows the $(K^{*0}+\overline{K^{*0}})/2$ and
$(K^{*+}+K^{*-})/2$ invariant yields for $|y| < $ 0.5 as a
function of $p_T$. The dotted curve in this figure is the fit to
the power-law function:
\begin{eqnarray}
\frac{1}{2\pi p_T}\frac{d^2N}{dydp_T} = \frac{dN}{dy} \frac{2
(n-1)
(n-2)}{\pi (n-3)^2 \langle p_T \rangle^2}  \nonumber \\
\left(1+\frac{p_T}{\langle p_T \rangle(n-3)/2}\right)^{-n},
\label{eq:power-law}
\end{eqnarray}
where $n$ is the order of the power law and $\langle p_T \rangle$
is the average transverse momentum. The data were fit for $p_T
> 0.5$ GeV/$c$. The power-law fit does not reproduce
the two first $p_T$ bins (0.0 $\leq p_T <$ 0.2 GeV/$c$ and 0.2
$\leq p_T <$ 0.4 GeV/$c$) since at low $p_T$ particle production
may be dominated by soft processes. From the power-law fit, the
$\chi^2/ndf$ is 0.93. The dashed curve in Fig.~\ref{fig:pT} is the
$K^{*0}$ spectrum fit to the exponential function from
Eq.~\ref{eq:exponential} and then extrapolated to higher $p_T$.
The data could not be described by this exponential fit indicating
that hard processes dominate the particle production for $p_T >
1.5$ GeV/$c$. Some model~\cite{levy1} suggests to use the Levy
function in Equation~\ref{eq:levy} to represent the $p_T$
spectrum:
\begin{eqnarray}
\frac{1}{2\pi p_T}\frac{d^2N}{dydp_T} = \frac{dN}{dy} \frac{(n-1)
(n-2)}{2\pi nT(nT+m_0(n-2))}  \nonumber \\
\left(1+\frac{\sqrt{p_T^2+m_0^2}-m_0}{nT}\right)^{-n}.
\label{eq:levy}
\end{eqnarray}
The dashed-dotted curve in Fig.~\ref{fig:pT} is the Levy function
fit with $\chi^2/ndf=$ 0.90 to the $K^*$ spectrum in all the
measured $p_T$ range ($p_T < 4$ GeV/$c$).

\subsection{Average Transverse Momentum $\langle p_T \rangle$}
In Au+Au collisions, the $p_T$ range of the exponential fit covers
$>$85\% of all the $K^*$ yield so that the $K^*$ average
transverse momentum ($\langle p_T \rangle$) can be reasonably
calculated by using the inverse slope parameter ($T$) extracted
from the exponential fit function and assuming the exponential
behavior over all the $p_T$ range:
\begin{equation}
\langle p_T \rangle = \frac{\int^\infty_0
p_T^2e^{-(\sqrt{p_T^2+m_0^2}-m_0)/T}dp_T \nonumber}{\int^\infty_0
p_Te^{-(\sqrt{p_T^2+m_0^2}-m_0)/T}dp_T}. \label{eq:mean}
\end{equation}

In $p+p$ collisions, the neutral and charged $K^*$ spectrum shown
in Fig.~\ref{fig:pT} covers $>$98\% of all the $K^*$ yield so that
the $\langle p_T \rangle$ is directly calculated from the data
points in the spectrum. The systematic uncertainty in $p+p$
includes the differences between this calculation and the
exponential fit to the $K^{*0}$ only at $p_T<1.6$ GeV/$c$, the
power-law fit to both neutral and charged $K^*$ at $p_T>1.5$
GeV/$c$, the Levy function fit at $p_T<4$ GeV/c. The systematic
uncertainties for all the $\langle p_T \rangle$ values include the
effects discussed in the previous section and the differences
caused by different fit functions to the invariant yield, such as
the Boltzmann fit ($m_T e^{-(m_T-m_0)/T}$) and the blast wave
model fit~\cite{blastwave}. The calculated $K^*$ $\langle p_T
\rangle$ for different centralities in Au+Au and minimum bias
$p+p$ collisions are listed in Table~\ref{tab:mean}.

\begin{table}[h]
\caption{\label{tab:mean}The $K^*$ $\langle p_{T} \rangle$ for
different centralities in Au+Au
  and minimum bias $p+p$ collisions. The first error is statistical, the second is systematic.}
\begin{ruledtabular}
\begin{tabular}{cc}
 &$\langle p_T \rangle$ (GeV/$c$)\\ \hline

    top 10\% central      & 1.08$\pm$0.03$\pm$0.12   \\
    0-10\%      & 1.08$\pm$0.08$\pm$0.12   \\
    10-30\%   & 1.12$\pm$0.06$\pm$0.13   \\
    30-50\%   & 1.08$\pm$0.05$\pm$0.12   \\
    50-80\%   & 1.03$\pm$0.04$\pm$0.12   \\
    $p+p$         & 0.81$\pm$0.02$\pm$0.14   \\
\end{tabular}
\end{ruledtabular}
\end{table}

\begin{figure}[htp]
\centering
\includegraphics[height=18pc,width=18pc]{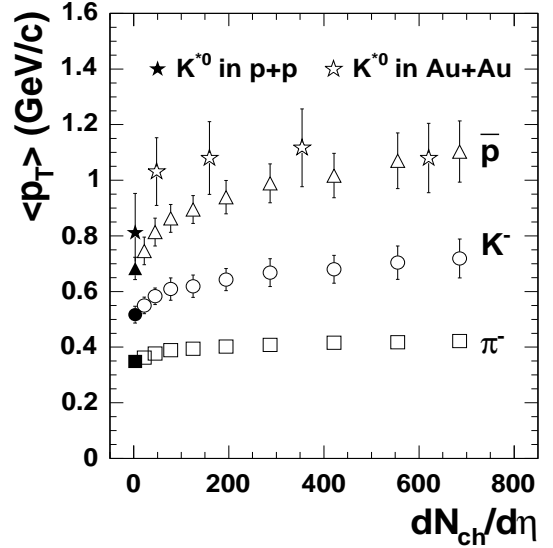}
\caption{The $K^{*}$ $\langle p_T \rangle$ as a function of
$dN_{ch}/d\eta$ compared to that of $\pi^-$, $K^-$, and
$\overline{p}$ for minimum bias $p+p$ (solid symbols) and Au+Au
(open symbols) collisions. The errors shown are the quadratic sum
of the statistical and systematic uncertainties.}\label{fig:mean}
\end{figure}

The $K^{*}$ $\langle p_T \rangle$ as a function of the charged
particle multiplicity ($dN_{ch}/d\eta$) is shown in
Fig.~\ref{fig:mean} and compared to that of $\pi^-$, $K^-$, and
$\overline{p}$~\cite{meanpt} for different centralities in Au+Au
and minimum bias $p+p$ collisions. The $K^{*0}$ $\langle
p_T\rangle$ in Au+Au collisions is significantly larger than in
minimum bias $p+p$ collisions. No significant centrality
dependence of $\langle p_T \rangle$ is observed for $K^*$ in Au+Au
collisions.

\subsection{Particle Ratios}
The $K^*$ vector meson and its corresponding ground state, the
$K$, have identical quark content in the context of the standard
model of particles. They differ only in their masses and the
relative orientation of their quark spins. Thus, the $K^*/K$ yield
ratio may be the most interesting and the least model dependent
ratio for studying the $K^*$ production properties and the
freeze-out conditions in relativistic heavy-ion collisions. The
$K^*$ and $\phi$ mesons have a very small mass difference, their
total spin difference is $\Delta S=0$, and both are vector mesons.
One significant difference between the $K^*$ and $\phi$ is their
lifetimes, with the $\phi$ meson lifetime being a factor of 10
longer than that of the $K^*$. Therefore, it is important to
measure the $\phi/K^*$ yield ratio and compare the potential
differences in $K^*/K$ and $\phi/K$ yield ratios in relativistic
heavy-ion collisions to study different hadronic interaction
effects on different resonances.

\begin{figure}[htp]
\centering
\includegraphics[height=22pc,width=18pc]{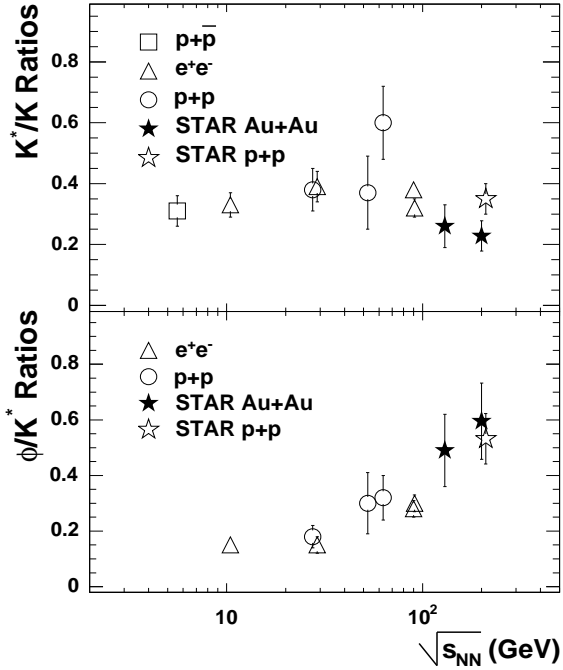}
\caption{The $K^{*}/K$ (upper panel) and $\phi/K^{*}$ (lower
panel) yield ratios as a function of the c.m. system energies. The
yield ratios for central Au+Au collisions at $\sqrt{s_{_{NN}}} = $
130~\cite{kstar130} and 200 GeV and minimum bias $p+p$
interactions at $\sqrt{s_{_{NN}}} = $ 200 GeV are compared to
measurements from e$^+$e$^-$ at $\sqrt{s}$ of 10.45 GeV
\cite{alb}, 29 GeV \cite{der} and 91 GeV \cite{abe,pei},
$\bar{p}p$ at $\sqrt{s}$ of 5.6 GeV \cite{can} and $pp$ at
$\sqrt{s}$ of 27.5 GeV \cite{agu}, 52.5 GeV \cite{dri} and 63 GeV
\cite{ake}. The errors at $\sqrt{s_{NN}}$= 130 and 200 GeV
correspond to the quadratic sum of the statistical and systematic
errors.}\label{fig:ratio_energy}
\end{figure}

The $K^{*}/K$ yield ratios as a function of the c.m. system
energies are shown in the upper panel of
Fig.~\ref{fig:ratio_energy}. The $K^{*}/K^-$ yield ratios for
central Au+Au collisions at $\sqrt{s_{_{NN}}} = $
130~\cite{kstar130} and 200 GeV and minimum bias $p+p$
interactions at $\sqrt{s_{_{NN}}} = $ 200 GeV are compared to
measurements in $e^+e^-$ \cite{alb,der,abe,pei}, $\overline{p}+p$
\cite{can}, and $p+p$ \cite{agu,dri,ake}. The $K^{*}/K^-$ yield
ratios depicted in Fig.~\ref{fig:ratio_energy} do not show a
strong dependence on the colliding system or the c.m. system
energy, with the exception of the $K^{*}/K^-$ yield ratio at
$\sqrt{s_{_{NN}}} = $ 200 GeV. In this case, the $K^{*}/K^-$ yield
ratio for central Au+Au collisions is significantly lower than the
minimum bias $p+p$ measurement at the same c.m. system energy. The
$\phi/K^{*}$ yield ratios as a function of the c.m. system
energies are depicted in the lower panel of
Fig.~\ref{fig:ratio_energy}. The $\phi/K^{*}$ yield ratios for
central Au+Au collisions at $\sqrt{s_{_{NN}}} = $
130~\cite{kstar130} and 200 GeV and minimum bias $p+p$
interactions at $\sqrt{s_{_{NN}}} = $ 200 GeV are compared to
measurements in $e^+e^-$ \cite{alb,der,abe,pei} and $p+p$
\cite{agu,dri,ake}. Figure \ref{fig:ratio_energy} shows an
increase of the yield ratio $\phi/K^{*}$ measured in Au+Au
collisions compared to the measurements in $p+p$ and $e^+e^-$ at
lower energies.

\begin{table}[h]
\caption{\label{tab:ratio}The $K^*/K^-$, $\phi/K^*$, and
$\phi/K^-$ yield ratios for different
  centralities in Au+Au and for minimum bias $p+p$ interactions. The first error is statistical, the second is systematic.}
\begin{ruledtabular}

\begin{tabular}{cccc}
 &$K^*/K$&$\phi/K^*$&$\phi/K$\\ \hline
    0-5\%     &                   &               &  0.16$\pm$0.01$\pm$0.02 \\
    0-10\%    & 0.23$\pm$0.01$\pm$0.05 & 0.60$\pm$0.06$\pm$0.12 & 0.15$\pm$0.01$\pm$0.02 \\
    10-30\%   & 0.24$\pm$0.02$\pm$0.05 & 0.63$\pm$0.07$\pm$0.14 & 0.16$\pm$0.01$\pm$0.02 \\
    30-50\%   & 0.26$\pm$0.02$\pm$0.06 & 0.58$\pm$0.06$\pm$0.13 & 0.16$\pm$0.01$\pm$0.02 \\
    50-80\%   & 0.26$\pm$0.02$\pm$0.05 & 0.53$\pm$0.05$\pm$0.11 & 0.15$\pm$0.01$\pm$0.02 \\
    $p+p$       & 0.35$\pm$0.01$\pm$0.05 & 0.53$\pm$0.03$\pm$0.09 & 0.14$\pm$0.01$\pm$0.02 \\
\end{tabular}
\end{ruledtabular}
\end{table}

\begin{figure}[htp]
\centering
\includegraphics[height=13pc,width=18pc]{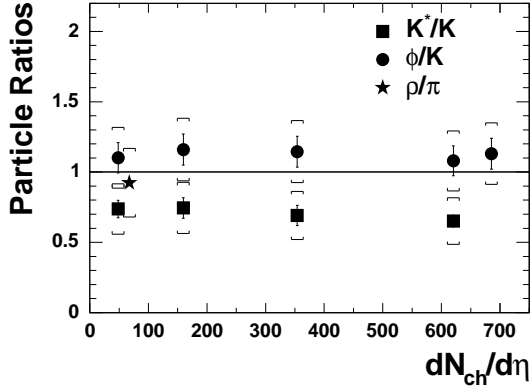}
\caption{The $K^{*}/K^-$, $\phi/K^-$, and $\rho^0/\pi^-$ yield
ratios as a function of $dN_{ch}/d\eta$ for Au+Au collisions at
$\sqrt{s_{_{NN}}}\!=\!$ 200 GeV. All yield ratios have been
normalized to the corresponding yield ratio measured in minimum
bias $p+p$ collisions at the same c.m. system energy and indicated
by the solid line. Both statistical and systematic uncertainties
are shown.}\label{fig:ratio}
\end{figure}

Table~\ref{tab:ratio} lists the $K^*/K^-$, $\phi/K^*$, and
$\phi/K^-$ yield ratios for different centralities in Au+Au and
minimum bias $p+p$ interactions. Figure~\ref{fig:ratio} depicts
the $K^*/K^-$, $\phi/K^-$~\cite{phi2}, and
$\rho^0/\pi^-$~\cite{rho1} yield ratios as a function of
$dN_{ch}/d\eta$ at $\sqrt{s_{_{NN}}} = $ 200 GeV. All yield ratios
have been normalized to the corresponding yield ratio measured in
minimum bias $p+p$ collisions at the same $\sqrt{s_{_{NN}}}$ and
indicated by the solid line in Fig.~\ref{fig:ratio}. As mentioned
previously and shown in Fig.~\ref{fig:ratio_energy}, the
$K^{*0}/K^-$ yield ratio for central Au+Au collisions is
significantly lower than the minimum bias $p+p$ measurement at the
same c.m. system energy. In addition, statistical model prediction
of $K^*/K$ of 0.33$\pm$0.01~\cite{rapp4,bron,pbm2} is considerably
larger (in a 2$\sigma$ effect) than than our measurement of
0.23$\pm$0.05 in 0-10\% Au+Au. The $K^{*0}$ regeneration depends
on $\sigma_{K\pi}$ while the rescattering of the daughter
particles depends on $\sigma_{\pi\pi}$ and $\sigma_{\pi p}$, which
are considerably larger (factor $\sim$5) than $\sigma_{K\pi}$
\cite{proto,matison}. The lower $K^{*0}/K^-$ yield ratio measured
may be due to the rescattering of the $K^{*0}$ decay products. The
$\rho^0/\pi^-$ yield ratio from minimum bias $p+p$ and peripheral
Au+Au interactions at the same c.m. system energy are comparable.
Due to the relatively long lifetime of the $\phi$ meson and the
negligible $\sigma_{KK}$, the rescattering of the $\phi$ decay
products and the $\phi$ regeneration should be negligible. The
statistical model calculations~\cite{bron,pbm2} predict the
$\phi/\pi^-$ yield ratio to be 0.025$\pm$0.001 while STAR measured
the $K^-/\pi^-$ yield ratio to be 0.15$\pm$0.02~\cite{meanpt}.
Thus the $\phi/K^-$ yield ratio combining the model prediction and
experimental measurements is 0.17$\pm$0.02 which successfully
reproduces the $\phi/K^-$ yield ratio measurement depicted in
Table~\ref{tab:ratio} and Fig.~\ref{fig:ratio}.

The centrality dependence of the resonance yield ratios depicted
in Fig.~\ref{fig:ratio} suggests that the $\phi$ regeneration and
the rescattering of the $\phi$ decay products are negligible, and
the rescattering of the $K^{*0}$ decay products is dominant over
the $K^{*0}$ regeneration and therefore the reaction channel $K^*
\leftrightarrow K\pi$ is not in balance. As a result, the
$K^{*0}/K^-$ yield ratio can be used to estimate the time between
chemical and kinetic freeze-outs:
\begin{equation}
\frac{K^*}{K}|_{\text{kinetic}}=\frac{K^*}{K}|_{\text{chemical}}
\times e^{-\Delta t/\tau}, \label{eq:time}
\end{equation}
where $\tau$ is the $K^*$ lifetime of 4 fm/$c$ and $\Delta t$ is
the time between chemical and kinetic freeze-outs. If we use the
minimum bias $p+p$ measurement of the $K^{*0}/K^-$ yield ratio as
the one at chemical freeze-out and use the most central
measurement of the $K^{*0}/K^-$ yield ratio in Au+Au collisions
for the production at kinetic freeze-out, then under the
assumptions that i) all the $K^*$s which decay before kinetic
freeze-out are lost due to the rescattering effect and ii) there's
no regeneration effect, the time between chemical and kinetic
freeze-outs is short and $\Delta t$ = 2 $\pm$ 1 fm/$c$. All the
above assumptions reduce the estimated $\Delta t$. Thus the
previous value is a lower limit of $\Delta t$ and it is not in
conflict with the estimations ($>$6 fm/$c$) in~\cite{meanpt}.
These two measurements together indicate that considerable
resonance regeneration effect may happen even (about 4 fm/$c$)
after chemical freeze-out.

\subsection{Elliptic Anisotropy $v_2$}
In non-central Au+Au collisions, the elliptic flow ($v_2$) is
defined as the second harmonic coefficient of the Fourier
expansion of the azimuthal particle distributions in momentum
space~\cite{art}. The $K^{*0}$ $v_2$ can be calculated as:
\begin{equation}
v_2=\langle \text{cos}[2(\phi-\Psi_r)] \rangle, \label{eq:v2}
\end{equation}
where $\phi$ is the $K^{*0}$ azimuthal angle in the momentum
space, $\Psi_r$ denotes the actual reaction plane angle and
$\langle\rangle$ indicates the average over all $K^{*0}$ in all
events.

For each $K\pi$ pair, the reaction plane angle was estimated by
the event plane ($\Psi_2$) which in turn was determined by using
all the primary tracks except the kaon and pion tracks in the
pair:
\begin{eqnarray}
\Psi_2=\frac{1}{2} \times \tan^{-1} \nonumber \\
\left(\frac{\sum_{i}^{} \omega_{i}\sin(2\phi_i)
-\omega_K\sin(2\phi_K)-\omega_\pi\sin(2\phi_\pi)}{\sum_i
\omega_{i}\cos(2\phi_i)-\omega_K\cos(2\phi_K)-\omega_\pi\cos(2\phi_\pi)}\right),
\end{eqnarray}
where $\omega_i$ is the weight for each track used to optimize the
event plane resolution, the subscripts $K$ and $\pi$ stand for the
kaon and pion candidate track, respectively. This prevents the
auto correlation between the $K\pi$ azimuthal angle $\phi_{K\pi}$
and the event plane angle $\Psi_2$~\cite{haibin}.

\begin{figure}[htp]
\centering
\includegraphics[height=13pc,width=18pc]{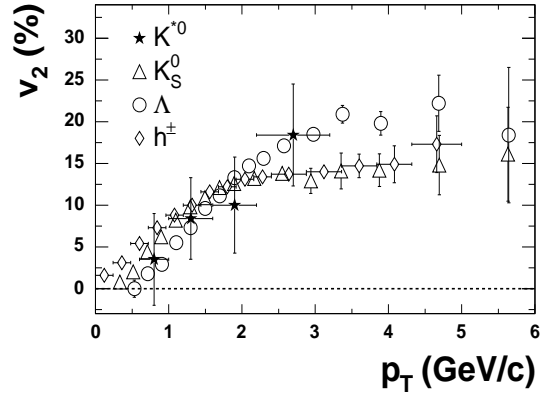}
\caption{The $K^{*0}$ $v_2$ (filled stars) as a function of $p_T$
for minimum bias Au+Au collisions compared to the $K_S^0$ (open
triangles), $\Lambda$ (open circles), and charged hadron (open
diamonds) $v_2$. The errors shown are statistical
only.}\label{fig:v2}
\end{figure}

In minimum bias Au+Au collisions, the unlike-sign and mixed-event
$K\pi$ pair invariant mass distributions are reconstructed in
$\text{cos}[2(\phi-\Psi_2)]$ bins and in $p_T$ bins. After the
mixed-event background subtraction for each
$\text{cos}[2(\phi-\Psi_2)]$ bin and $p_T$ bin, the $K^{*0}$
yields are then obtained as a function of
$\text{cos}[2(\phi-\Psi_2)]$ for given $p_T$ bin. The average
$\langle \text{cos}[2(\phi-\Psi_2)] \rangle$ is then calculated
for each $p_T$ bin. The finite resolution of the event plane
angle, which is due to the limited number of tracks in the event
plane calculation, reduces the measured $K^{*0}$ $v_2$. Thus the
above obtained $\langle \text{cos}[2(\phi-\Psi_2)] \rangle$ values
are further corrected for an event plane resolution factor ($<1$)
using the method presented in~\cite{art}. Figure~\ref{fig:v2}
shows the $K^{*0}$ $v_2$ as a function of $p_T$ compared to the
$K_S^0$, $\Lambda$, and charged hadron $v_2$ for minimum bias
Au+Au collisions~\cite{KsLav2}. A significant non-zero $K^{*0}$
$v_2$ is observed. Nevertheless, due to the large uncertainties on
the $K^{*0}$ $v_2$ measurement, no significant difference is
observed between the $K^{*0}$ $v_2$ and the $K_S^0$, $\Lambda$,
and charged hadron $v_2$.

In order to calculate the contributions to the $K^*$ production
from either direct quark or hadron combinations, the following
function~\cite{dong} was used to fit the $K^{*0}$ $v_2$:
\begin{equation}
v_2(p_T,n)=\frac{an}{1+\text{exp}[-(p_T/n-b)/c]}-dn,
\label{eq:v2_scaling}
\end{equation}
where $a$, $b$, $c$ and $d$ are constants extracted by fitting to
the $K_S^0$ and $\Lambda$ $v_2$ data points in~\cite{dong}, and
$n$ is the open parameter standing for the number of constituent
quarks. From the fit to the $K^{*0}$ $v_2$, $n=3\pm 2$ was
obtained. Due to the large statistical uncertainties, it is
difficult to identify the $K^*$ production fractions from direct
quark combinations ($n=2$) or hadron combinations ($n=4$). About
15-20 times more Au+Au collision events were taken by the STAR
experiment in the fourth RHIC run in 2004 which is expected to
provide enough sensitivity for more precise calculations of $n$ to
identify the $K^*$ from different production mechanism.

\subsection{Nuclear Modification Factor}
The number of binary collisions ($N_{bin}$) scaled centrality
ratio ($R_{CP}$) is a measure of the particle production
dependence on the size and density of the collision system and is
closely related to the nuclear modification factor ($R_{AA}$).
Recent measurements of the $\Lambda$ and $K_S^0$ $R_{CP}$ at
RHIC~\cite{KsLav2} have shown that in the intermediate $p_T$
region ($2 < p_T < 4$ GeV/$c$), the $\Lambda$ and $K_S^0$ $R_{CP}$
are significantly smaller than unity. These measurements suggest
that high $p_T$ jets lose energy through gluon radiation while
traversing through dense matter. It has also been observed that
the $R_{CP}$ is significantly different for $\Lambda$ and $K_S^0$
with $p_T > 2$ GeV/$c$. It is not clear whether this $R_{CP}$
difference is due to a mass or a particle species effect. The
$K^*$ is a meson but has a mass that is close to the $\Lambda$
baryon mass. Thus, the measurement of the $K^*$ $R_{CP}$ may help
in discriminating between mass or particle species effect at the
intermediate $p_T$ region.

The $K^*$ $R_{CP}$ was obtained from the $p_T$ spectra of the top
10\% and the 50-80\% most peripheral Au+Au collisions. The $K^*$
$R_{AA}$ was calculated from the $p_T$ spectrum of the 10\% most
central Au+Au collisions and the $p_T$ spectrum of the minimum
bias $p+p$ collisions.

\begin{figure}[htp]
\centering
\includegraphics[height=15pc,width=18pc]{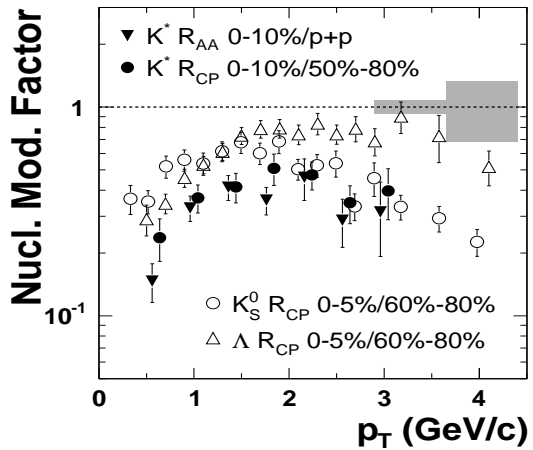}
\caption{The $K^{*}$ $R_{AA}$ (filled triangles) and $R_{CP}$
(filled circles) as a function of $p_T$ compared to the $K_S^0$
(open circles) and $\Lambda$ (open triangles) $R_{CP}$. The errors
shown are statistical only. The dashed line represents the number
of binary collisions scaling. The widths of the grey bands
represent the systematic uncertainties of $R_{AA}$ (left) and
$R_{CP}$ (right) due to the model calculations of
$N_{bin}$.}\label{fig:raa}
\end{figure}

The $K^*$ $R_{AA}$ and $R_{CP}$ as a function of $p_T$ compared to
the $\Lambda$ and $K_S^0$ $R_{CP}$ are shown in
Fig.~\ref{fig:raa}. The $K^*$ $R_{AA}$ and $R_{CP}$ for $p_T <
1.6$ GeV/$c$ are smaller than the $\Lambda$ and $K_S^0$ $R_{CP}$
indicating the strong rescattering of the $K^*$ daughters at low
$p_T$. The rescattering of the $K^*$ decay products is weaker for
$p_T > 1.6$ GeV/$c$ since $K^*$ with larger $p_T$ are more likely
to decay outside the fireball~\cite{bleicher0}. Therefore, larger
$p_T$ $K^*$ have a larger probability to be measured compared to
low $p_T$ $K^*$. The $K^*$ $R_{AA}$ and $R_{CP}$ are closer to the
$K_S^0$ $R_{CP}$ and different from the $\Lambda$ $R_{CP}$ for
$p_T > 1.6$ GeV/$c$. Thus, a strong mass dependence of the nuclear
modification factor is not supported and a baryon-meson effect is
favored in the particle production in the intermediate $p_T$
region.

\section{Conclusion}
Results on the $K^{*0}$ and $K^{*\pm}$ resonance production in
Au+Au and $p+p$ collisions measured with the STAR experiment at
$\sqrt{s_{_{NN}}}$ = 200 GeV were presented. The $K^{*0}$ and
$K^{*\pm}$ signals were reconstructed via their hadronic decay
channels $K^{*0}\rightarrow K\pi$ and $K^{*\pm}\rightarrow
K_S^0\pi^{\pm}$ at midrapidity.

The $K^{*0}/K$ yield ratios in Au+Au collisions were observed to
be smaller than the ratio in $p+p$ interactions which may be
interpreted in the context of finite cross sections in a late
hadronic phase. The result suggests that the rescattering of the
$K^{*0}$ decay products is dominant over the $K^{*0}$ regeneration
and therefore the reaction channel $K^* \leftrightarrow K\pi$ is
not in balance. As a result, the $K^{*0}/K^-$ yield ratio can be
used to estimate the time between chemical and kinetic
freeze-outs. Using the $K^{*0}/K^-$ yield ratio, the lower limit
of the time between chemical and kinetic freeze-outs is estimated
to be at least $2\pm1$ fm/$c$.

A significant non-zero $K^{*0}$ elliptic flow $v_2$ was measured
as a function of $p_T$ in minimum bias Au+Au collisions. Due to
limited statistics, no conclusive statement can be made about the
difference between the $K^{*0}$ $v_2$ and the $K_S^0$, $\Lambda$,
and charged hadron $v_2$. The estimated number of constituent
quarks for the $K^{*0}$ from the $v_2$ scaling according to
Equation~\ref{eq:v2_scaling} is $3\pm 2$. Thus, larger statistics
for Au+Au collision data are needed to identify the $K^*$
production fractions from direct quark combinations or hadron
combinations.

The $K^{*0}$ nuclear modification factors $R_{AA}$ and $R_{CP}$
were measured as a function of $p_T$. Both the $K^{*0}$ $R_{AA}$
and $R_{CP}$ are found to be closer to the $K_S^0$ $R_{CP}$ and
different from the $\Lambda$ $R_{CP}$ for $p_T > 2$ GeV/$c$. A
strong mass dependence of the nuclear modification factor is not
observed. This establishes a baryon-meson effect over a mass
effect in the particle production at the intermediate $p_T$
region.

\section{Acknowledgement}
We thank the RHIC Operations Group and RCF at BNL, and the NERSC
Center at LBNL for their support. This work was supported in part
by the HENP Divisions of the Office of Science of the U.S. DOE;
the U.S. NSF; the BMBF of Germany; IN2P3, RA, RPL, and EMN of
France; EPSRC of the United Kingdom; FAPESP of Brazil; the Russian
Ministry of Science and Technology; the Ministry of Education and
the NNSFC of China; Grant Agency of the Czech Republic, FOM and UU
of the Netherlands, DAE, DST, and CSIR of the Government of India;
Swiss NSF; and the Polish State Committee for Scientific Research.

\end{document}